# smFISH_batchRun: A smFISH image processing tool for single-molecule RNA Detection and 3D reconstruction

Nimmy S. John[1,2], and ChangHwan Lee[1,2,3]

[1]Department of Biological Sciences, University at Albany, New York, NY, USA

[2]The RNA Institute, University at Albany, New York, NY, USA

[3]Corresponding author

## Abstract

Single-molecule RNA imaging has been made possible with the recent advances in microscopy methods. However, systematic analysis of these images has been challenging due to the highly variable background noise, even after applying sophisticated computational clearing methods. Here, we describe our custom MATLAB scripts that allow us to detect both nuclear nascent transcripts at the active transcription sites (ATS) and mature cytoplasmic mRNAs with single-molecule precision and reconstruct the tissue in 3D for further analysis. Our codes were initially optimized for the *C. elegans* germline but were designed to be broadly applicable to other species and tissue types.

## Keywords

MATLAB source code, Fluorescence microscopy, Image processing, smFISH, RNA Detection, *C. elegans*

| Code metadata | |
|---|---|
| Current code version | 2.0 |
| Permanent link to code/repository used for this code version | https://github.com/chleelab/smFISH_detection |
| Permanent link to reproducible capsule | N/A |
| Legal code license | MIT License |
| Code versioning system used | None |
| Software code languages, tools and services used | MATLAB |
| Compilation requirements, operating environments and dependencies | None |
| If available, link to developer documentation/manual | N/A |
| Support email for questions | chlee@albany.edu |

## 1. Introduction

Recent advances in microscopy, particularly single-molecule Fluorescent *In Situ* Hybridization (smFISH), have enabled the imaging of biomolecules at single-molecule resolution[1–6]. These techniques provide powerful tools to investigate the mechanistic basis of biological processes and their regulations in various *in vitro* and *in vivo* systems, including mammalian tissue culture,

*C. elegans, Drosophila*, and mouse[3,5,7,8]. smFISH visualizes both nascent transcripts at the active transcription sites (ATS) and mature mRNAs in the cytoplasm, by targeting the intronic and exonic regions with spectrally distinct fluorophores[3–7]. This approach provides direct transcriptional readouts for the gene's transcriptional activity and its spatial regulation[3].

smFISH has been widely used to locate and quantify key components and targets of various signaling pathways, including Notch, Wnt, and TGF-β[1–3,5,9]. However, many studies have relied on either simple manual counting or smFISH image analysis tools like FISH-quant[10,11], Radial Symmetry FISH (RS-FISH)[12], and FISHtoFigure[13], which are primarily optimized for flat tissue culture systems and offer limited flexibility for tailoring analyses to specific sample or tissue types, constraining the depth and precision of smFISH analysis.

Here, we present our custom MATLAB scripts for automated smFISH image analysis, enabling the detection of different RNA species and subcellular organelles (e.g., nuclei, mitochondria, nucleolus), 3D reconstruction of the tissues, and their spatial pattern analysis. Originally developed for recording ATS, mRNA, nuclei, and protein signals in the *C. elegans* germline, these scripts are adaptable to a wide range of tissue types and species.

## 2. Description

**Image acquisition settings for smFISH**

Essentially worms at any developmental stage can be used for smFISH analysis via our custom scripts. To validate their performance, synchronized L1 larvae[14] were grown on OP50 until day 1 of adulthood, dissected, and prepped for smFISH as previously described[3,5,6]. smFISH for *sygl-1* was performed on the prepared samples as previously described[3–6]. Gonads were also imaged using a Leica DMi8 Widefield Microscope equipped with the THUNDER Imaging system with computational image clearing methods to remove excessive background as previously described[5,6]. All gonads were imaged completely (depth >15 μm) with a Z-step size of 0.3 μm using the Leica Application Suite X (LAS X) acquisition software (Leica Microsystems Inc., Buffalo Grove, IL) and with LED8 light sources as described previously[6].

**Image preparation for subsequent RNA Detection**

For systematic spatial RNA analysis, *C. elegans* germline images need to be oriented such that the distal end is aligned perpendicular to the left-hand side of the image. We used Fiji (ImageJ) to rotate and reposition the images accordingly. The image can be cropped for faster processing. We cropped the distal 60-65 μm of the gonads and then saved the output in the lossless TIFF format. Cropped images with identical experimental conditions can be saved in one folder and processed in a batch to streamline processing.

**RNA Detection via custom MATLAB scripts**

The custom MATLAB script for RNA detection builds on our previous work and has been further optimized for more precise RNA and nuclear detection across a range of imaging platforms, including wide-field, confocal, and super-resolution microscopy. These improvements include newly implemented functions, additional parameters, and cross-validation steps to improve object detection accuracy and compatibility of a broader range of cellular systems[3,5,8,15]. It also automates threshold optimization for RNA detection by systematically applying and evaluating multiple thresholds to identify the most effective settings for accurate image processing and analysis.

The MATLAB script, "smFISH_batchRun", includes user-defined variables to specify input and output paths for smFISH image processing. Once these paths are defined, the script enables automated processing using multiple threshold values to accommodate variability in signal-to-background ratios across different images and samples. The variables "ExonThresRange" and "IntronThresRange" in this script can be specified as single threshold values or as ranges with desired intervals, allowing flexible thresholding for exon and intron signal detection. Since the image channels are sequentially acquired in decreasing wavelength order, the variable "ChOrder" allows the users to define the identity of each channel, where '1' corresponds to the nucleus channel, '2' to exon, '3' to intron, '4' to immunostaining (e.g., protein staining), '5' to the distal tip cell (DTC) channel, and '0' to skip a channel if it needs to be excluded from analysis or is not compatible with the detection code.

Once the main parameters are defined, the function "Batch_smFISH_v2_10" processes the prepped images using the specified intron and exon threshold values. This function constitutes the core of the detection pipeline. The script records and calculates several cellular features, such as nuclear radius, Voronoi cell boundaries, and the number of ATS or mRNAs in a cell, as well as the 3D coordinates of detected biomolecules, and their sizes and signal intensities. A key component of the workflow is the cross-validation between intron and exon channels to robustly identify ATS. All processed data are compiled into a structured output variable "af", with each type of measurement stored in its designated column for subsequent analysis.

Once the variable "af" is generated, users can access the accuracy of RNA and nuclear detection using two visualization functions: "VisualDNAdetect" for nuclear detection and "VisualRNAdetect" for ATS and mRNA detection. These functions allow users to visually evaluate how well the detection results match the image data for a given set of intron, exon, and nuclear thresholds, as illustrated in Figure 1. "VisualDNAdetect" requires only the "af" variable, whereas the "VisualRNAdetect" requires several input parameters: the "af" variable, RNA channel identity, a marker to represent detected spots in the image, and a brightness value to control the appearance of the MATLAB-generated output. The RNA channel number is defined according to the order specified in the "ChOrder" variable. For example, if the "ChOrder" is set to [3 2 1] (where '1' is for nucleus, '2' for exon, and '3' for intron), RNA channel 1 represents the intron/ATS information and the RNA channel 2 would be representing the mRNA information. If detection results appear under or overestimated, users can easily modify the intron and exon threshold in the main detection script, "smFISH_batchRun", and reprocess the images. Once optimal thresholds are established, the extracted data can be used for further quantitative and spatial analyses.

## 3. Impact

Since the advent of highly precise and direct methods to visualize RNA at the single-molecule level, analyzing the resulting images has remained a challenge due to low signal-to-background ratios, stemming from highly variable RNA signals and substantial background noise, particularly in wide-field microscopy. To our knowledge, smFISH_batchRun is the first automated script capable of accurately detecting RNA molecules in 3D, optimized for *C. elegans* tissues, which can also be used for other species, providing robust performance even in the presence of considerable background noise. This suite of scripts has enabled us to investigate the molecular consequences and mechanistic effects of physiological factors (e.g., aging, hypoxia) and genetic perturbations (e.g., pathological Notch mutations) on germline stem cell (GSC) regulation, demonstrating its broad applicability and flexibility across diverse experimental settings and conditions[5,6].

### 4. Limitations and potential improvements

These scripts were primarily optimized for *sygl-1* smFISH detection. We confirmed that other genes with varying expression levels and spatial patterns can also be detected with minor modifications (e.g., *lst-1, let-858*). However, broader applications, particularly for genes that are sparsely and weakly expressed, may require further optimization to ensure accurate and robust detection. The majority of smFISH image datasets used for training and optimization were derived from *C. elegans*. Expanding training to include image sets from diverse species and tissue types will enhance the applicability, flexibility, and robustness of our scripts across broader biological contexts.

### Acknowledgments

We are thankful for the resources provided by the Molecular Biology Core Facility in the Life Sciences Research Building and the RNA Institute at the University at Albany. This work was funded by the University at Albany (FRAP-A Award 1189585-1-97969).

### References

1. Raj, A. & van Oudenaarden, A. Single-molecule approaches to stochastic gene expression. *Annu Rev Biophys* **38**, 255–270 (2009).

2. Raj, A., van den Bogaard, P., Rifkin, S. A., van Oudenaarden, A. & Tyagi, S. Imaging individual mRNA molecules using multiple singly labeled probes. *Nature Methods* **5**, 877–879 (2008).

3. Lee, C., Sorensen, E. B., Lynch, T. R. & Kimble, J. C. elegans GLP-1/Notch activates transcription in a probability gradient across the germline stem cell pool. *eLife* **5**, e18370 (2016).

4. Lee, C. *et al.* Single-molecule RNA Fluorescence in situ Hybridization (smFISH) in Caenorhabditis elegans. *Bio-protocol* **7**, e2357–e2357 (2017).

5. Urman, M. A., John, N. S., Jung, T. & Lee, C. Aging disrupts spatiotemporal regulation of germline stem cells and niche integrity. *Biology Open* bio.060261 (2023) doi:10.1242/bio.060261.

6. John, N. S., Urman, M. A., Mehmood, M. G. & Lee, C. Genetic mutations in GLP-1/Notch pathway reveal distinct mechanisms of Notch signaling in germline stem cell regulation. 2025.03.25.645284 Preprint at https://doi.org/10.1101/2025.03.25.645284 (2025).

7. Crittenden, S. L. *et al.* Sexual dimorphism of niche architecture and regulation of the Caenorhabditis elegans germline stem cell pool. *Mol Biol Cell* **30**, 1757–1769 (2019).

8. Lynch, T. R., Xue, M., Czerniak, C. W., Lee, C. & Kimble, J. Notch-dependent DNA cis-regulatory elements and their dose-dependent control of C. elegans stem cell self-renewal. *Development* **149**, dev200332 (2022).

9. Wang, S. Single Molecule RNA FISH (smFISH) in Whole-Mount Mouse Embryonic Organs. *Current Protocols in Cell Biology* **83**, e79 (2019).

10. Mueller, F. *et al.* FISH-quant: automatic counting of transcripts in 3D FISH images. *Nat Methods* **10**, 277–278 (2013).

11. Imbert, A. *et al.* FISH-quant v2: a scalable and modular tool for smFISH image analysis. *RNA* **28**, 786–795 (2022).

12. Bahry, E. *et al.* RS-FISH: precise, interactive, fast, and scalable FISH spot detection. *Nat Methods* **19**, 1563–1567 (2022).

13. Bentley-Abbot, C. *et al.* An easy to use tool for the analysis of subcellular mRNA transcript colocalisation in smFISH data. *Sci Rep* **14**, 8348 (2024).

14. Kershner, A. M., Shin, H., Hansen, T. J. & Kimble, J. Discovery of two GLP-1/Notch target genes that account for the role of GLP-1/Notch signaling in stem cell maintenance. *PNAS* **111**, 3739–3744 (2014).

15. Crittenden, S. L. *et al.* Sexual dimorphism of niche architecture and regulation of the *Caenorhabditis elegans* germline stem cell pool. *MBoC* **30**, 1757–1769 (2019).

Figure 1

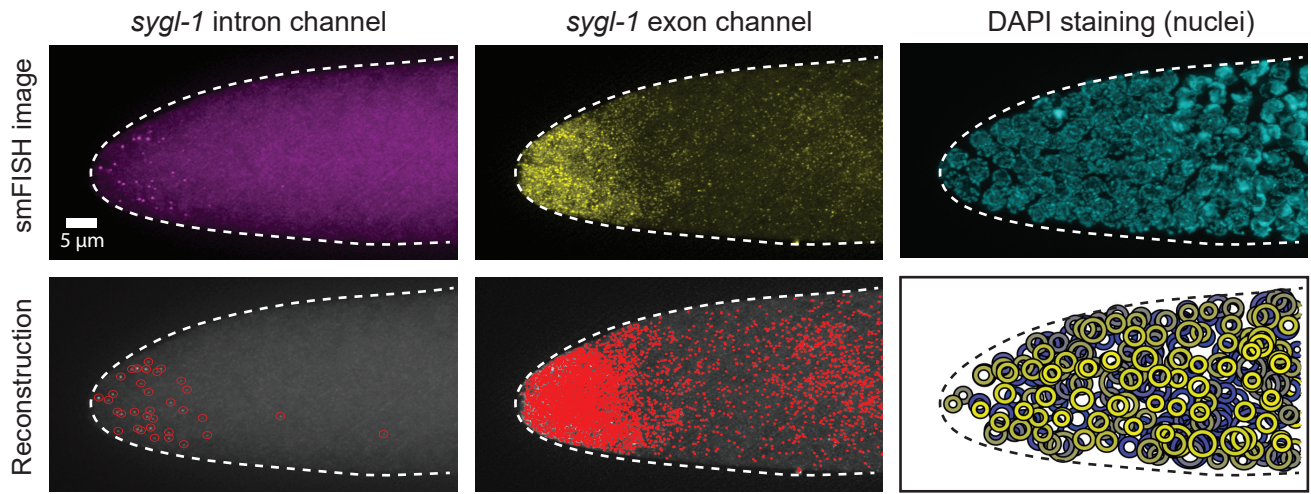